\documentclass[aps,prb,twocolumn,groupedaddress,showpacs,superscriptaddress,amssymb,amsmath,noeprint]{revtex4-2}
\usepackage{graphicx}
\usepackage[english]{babel}
\usepackage{dcolumn}
\usepackage{bm}
\usepackage{hyperref}
\usepackage{physics}
\usepackage{float}
\usepackage{cleveref}
\usepackage{multirow}
\usepackage{braket}
\usepackage{mathtools}
\usepackage{ulem}
\usepackage{soul}
\hypersetup{
    colorlinks=true,
    linkcolor=blue,
    urlcolor=blue,
    citecolor=blue
}
\usepackage{xcolor}
\usepackage{comment}

\newcommand{\dn}{\downarrow}

\newcommand{\GG}{\Gamma_G}
\newcommand{\Gd}{\Gamma_d}
\newcommand{\Cd}{C_{\rm diag}}
\newcommand{\In}[1]{\mathbb{I}_{#1}}

\newcommand{\sutdphys}{Science, Mathematics and Technology Cluster, Singapore
University of Technology and Design, 8 Somapah Road, 487372 Singapore}
\newcommand{\sutdepd}{EPD Pillar, Singapore University of Technology and Design, 8 Somapah Road, 487372 Singapore}

\newcommand{\cqt}{Centre for Quantum Technologies, National University of Singapore 117543, Singapore} 
\newcommand{\majulab}{MajuLab, CNRS-UNS-NUS-NTU International Joint Research Unit, UMI 3654, Singapore}  

\begin{document}

\title{Quantum dynamics in lattices in presence of bulk dephasing and a localized source}

\author{Tamoghna Ray}
\email{tamoghna.ray@icts.res.in}
\affiliation{International Centre for Theoretical Sciences, Tata Institute of Fundamental Research, Bangalore 560089, India}

\author{Katha Ganguly}
\email{katha.ganguly@students.iiserpune.ac.in}
\affiliation{Department of Physics, Indian Institute of Science Education and Research Pune, Dr. Homi Bhabha Road, Ward No. 8, NCL Colony, Pashan, Pune, Maharashtra 411008, India}

\author{Dario Poletti} 
\email{dario_poletti@sutd.edu.sg }
\affiliation{\sutdphys}
\affiliation{\sutdepd}
\affiliation{\cqt}
\affiliation{\majulab}

\author{Manas Kulkarni}
\email{manas.kulkarni@icts.res.in}
\affiliation{International Centre for Theoretical Sciences, Tata Institute of Fundamental Research, Bangalore 560089, India}

\author{Bijay Kumar Agarwalla}
\email{bijay@iiserpune.ac.in}
\affiliation{Department of Physics, Indian Institute of Science Education and Research Pune, Dr. Homi Bhabha Road, Ward No. 8, NCL Colony, Pashan, Pune, Maharashtra 411008, India}

\date{\today} 

\begin{abstract}
The aim of this work is to study the dynamics of quantum systems subjected to a localized fermionic source in the presence of bulk dephasing. We consider two classes of one-dimensional lattice systems: (i) a non-interacting lattice with nearest-neighbor and beyond, i.e., long-ranged (power-law) hopping, and (ii) a lattice that is interacting via short-range interactions modeled by a fermionic quartic Hamiltonian. We study the evolution of the local density profile $n_i(t)$ within the system and the growth of the total particle number $N(t)$ in it. For case (i), we provide analytical insights into the dynamics of the nearest-neighbor model using an adiabatic approximation, which relies on assuming faster relaxation of coherences of the single particle density matrix. For case (ii), we perform numerical computations using the time-evolving block decimation (TEBD) algorithm and analyze the density profile and the growth exponent in $N(t)$. Our detailed study reveals an interesting interplay between Hamiltonian dynamics and various environmentally induced mechanisms in open quantum systems, such as local source and bulk dephasing. It brings out rich dynamics, including universal dynamical scaling and anomalous behavior across various time scales and is of relevance to various quantum simulation platforms.
\end{abstract}

\maketitle

\section{Introduction}
Quantum dynamics of open many-body systems is of significant interest both from a theoretical~\cite{AMJFeb1982,MBPJan1998,BPOQS,IROct2015,HPBApr2016,ACApr2017,SGJul2017,TGLDec2022,MAP2025,PN2025,RS2025} and an experimental~\cite{AH2012,IB2012,RB2012,CM2019,ACJune2020,PZNov2024, XZJan2025} perspective. Many interesting features of the underlying setup often emerge while studying the spreading of local excitations or multi-time correlation functions~\cite{LBMar2015,JSBJan2018,ACApr2017, APMay2018,ASJan2020,PEDSep2020,ADAug2024,DSB2024,ADAug2024}, which are often directly linked to experimental observables~\cite{PR2014,MKJ2022}. This further helps to understand different dynamical regimes and characterize possible universality classes~\cite{JMDOct2003,MZApr2010,LBMar2015,RSJan2017,ML2017,DL2017,SGJul2017,APApr2019,MSJun2022,KGDec2024,YPW2024}. More precisely, the dynamics are often classified within diffusive, anomalous (sub-diffusive, super-diffusive), and ballistic universality classes. 
Such classification is often done via extracting the spatio-temporal exponents that appear in local excitation dynamics or dynamics of non-local observables such as number fluctuations in a domain~\cite{RSJan2017,DL2017,YPW2024,fluc_dephasing_sasamoto,PRLFS2025}. Alternatively, such a classification is also possible via system-size scaling dependence in 
non-equilibrium steady state quantum transport~\cite{JLDApr1990,MZApr2010,TPSep2012,APDec2021,MSJun2022,MSMay2023,MSOct2023,SSApr2024,KGDec2024}. 

In the context of open systems, subjecting a system to a localized source \cite{MB2010,PKNov2019,PKJun2020,ATNov2023} is a natural starting point, not only from the viewpoint of universality classes, but also to develop a thorough understanding of quantum dynamics across various time-scales, including the approach to equilibration. A complementary setup consisting of a localized loss (rather than a source)~\cite{PBOct2011,KVKJun2012,AMVJul2022,SUNov2022,SU2023,KGDec2024,KK2025,AC2022} is of significant experimental value~\cite{LCNov2019,MLNov2019,BBMA2020,MZHMay2023} given that many systems are often prone to inevitable particle losses. In the context of a localized source, subjected to an initially empty generic many-body lattice system, a natural question to ask is how the filling of particles and the local density profile evolve dynamically, before equilibration takes place. Such questions have been recently addressed in a variety of contexts for both bosonic~\cite{PKNov2019} and fermionic~\cite{PKJun2020,ATNov2023} non-interacting systems. However, the filling dynamics of lattice systems subjected to bulk (i.e., on all lattice sites) dephasing have remained unexplored. In such cases, one expects the emergence of interesting dynamics due to the interplay between unitary dynamics inherent to the lattice and dynamics induced by local source and bulk dephasing. It is important to note that the impact of such bulk dephasing mechanism in the context of steady-state transport has been explored extensively, where emergence of anomalous system-size scaling has been reported \cite{MZApr2010,ADAug2024,APDec2021,MSJun2022,MSMay2023,MSOct2023,SSApr2024,KGDec2024,LandiPRB2021}.

In this work, we fill a gap in the literature by studying the quantum dynamics of systems subjected to bulk dephasing and a localized source. We investigate two interesting classes of systems subjected to a dephasing mechanism at each of its lattice sites: (i) the lattice (either nearest-neighbor or long-ranged) is non-interacting and (ii) an inherently interacting short-ranged lattice. We study the quantum dynamics of particles filling the lattice, starting with vacuum initial conditions. Two primary quantities of interest are, (a) the evolution of the local density profile within the lattice and (b) the filling of quantum particles, i.e., the total number of particles injected into the system. The first setup is amenable to analytical calculations (in the nearest-neighbor case) that show how universality classes emerge in the asymptotic limit when the dynamics of total filling is considered. The second setup (inherently interacting case) is dealt with by employing the 
time-evolving block decimation (TEBD) algorithm~\cite{MZNov2004,USJan2011}.

We organize our paper as follows: In Sec.~\ref{sec:setupI}, we introduce first the short-ranged non-interacting setup with dephasing and present both numerical and detailed analytical results following the adiabatic approximation. In Sec.~\ref{sec:longr} we discuss our numerical findings of anomalous quantum dynamics in the long-ranged (power-law hopping) case. 
In Sec.~\ref{sec:inherently_interacting}, we provide results for the inherently interacting lattice setup subjected to dephasing following the TEBD approach. Finally, we summarize our results along with an outlook in Sec.~\ref{sec:summary}. Certain technical details are provided in appendices. 

\section{Tight-binding lattice with bulk dephasing and local source }
\label{sec:setupI}
\begin{figure}
    \centering
    \includegraphics[width=\linewidth]{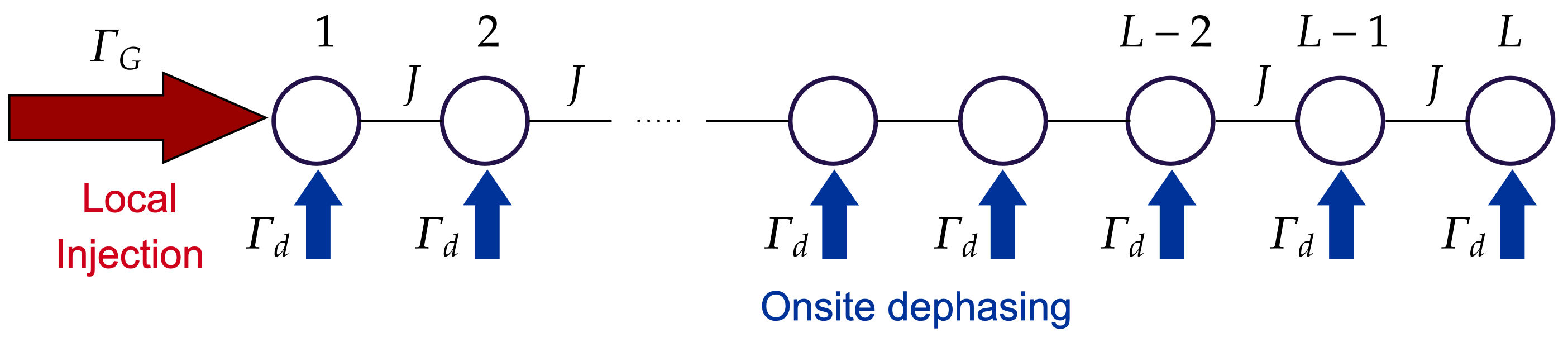}
    \caption{Schematic for nearest-neighbor lattice system of size $L$ with hopping strength $J$ subjected to onsite local dephasing with strength $\Gamma_d$ at each of its site. A local source is injecting quantum particles from the left end with rate $\Gamma_G$. The setup is modeled by Eq.~\eqref{eq:lindblad}.}
    \label{fig:schematic}
\end{figure}

We consider a nearest-neighbor tight-binding lattice that consists of fermions and is described by the Hamiltonian $H_S$ as
\begin{equation}
H_S=-J \sum_{i=1}^{L-1} c^{\dagger}_{i}c_{i+1}+{\rm h.c.},
\label{eq:system_hamiltonian_non_int}
\end{equation} 
where $J$ is the nearest-neighbor hopping amplitude and $c_i$ ($c_i^{\dagger})$ is the fermionic annihilation (creation) operator at the $i$-th site. 
The lattice is further subjected to dephasing at each lattice site. The schematic of our setup is given in Fig.~\ref{fig:schematic}. Such a dephasing mechanism is routinely employed to induce inelastic scattering and phase randomization processes \cite{MZMay2010,TGLDec2022,LY2024,SLJun2024}.
Here we focus on the dynamics of filling of fermions in such a dephased lattice system when the particles are injected from one end of the lattice. We model the dynamics of the system by a Gorini–Kossakowski–Sudarshan–Lindblad (GKSL) quantum master equation as~\cite{VG1976,L1976,BPOQS}, (we set $\hbar=1$ throughout the paper)
\begin{eqnarray}
    \dv{\rho}{t}&=&- {\rm i}[H_S,\rho]+\Gamma_G {\cal D}_G[\rho] + \Gamma_d {\cal D}_d[\rho],
    \label{eq:lindblad}
\end{eqnarray}
where the first term corresponds to the unitary dynamics induced by the lattice Hamiltonian. The second and third terms are the Lindblad dissipators. The dissipator ${\cal D}_{G}[\rho]$ is responsible for injecting particles at the first site of the lattice with rate $\Gamma_G$ and its form is given as 
\begin{equation}
{\cal D}_{G}[\rho] = c^{\dagger}_1 \,\rho \, c_1-\frac{1}{2}\{c_{1}c^{\dagger}_1,\rho\},
\end{equation}
and ${\cal D}_{d}[\rho]$ is the dissipator corresponding to the dephasing mechanism with rate $\Gamma_d$ and is given as
\begin{equation}
{\cal D}_{d}[\rho] = \sum_{i=1}^{L} \big[n_i,\big[n_i,\rho\big]\big],
\label{eq:lindblad-dephasing}
\end{equation}
where $n_i = c_i^\dagger c_i$ is the local occupation at site $i$. Given this setup, in what follows, we first provide the numerical results for the time dynamics of the local density profile and the growth of the total number of particles in the system. 

For the given model, one can study the dynamics of the system from the evolution of the two-point correlation matrix 
\begin{equation}
C_{i,j}(t) = {\rm Tr} \big[c^\dagger_i c_j \rho(t)\big], \quad i, j=1, 2, \cdots L.
\end{equation} 
The equation of motion of the two-point correlation function in the presence of a local source and bulk dephasing is given by
\begin{align}
    \dv{C_{i,j}}{t} =& - {\rm i} J \left( C_{i-1,j}+ C_{i+1,j} - C_{i,j-1} - C_{i,j+1} \right)\nonumber\\
    &- \frac{\Gamma_G}{2} (\delta_{1,i}+\delta_{1,j}) C_{i,j} +\Gamma_d\, (\delta_{i,j}-1) C_{i,j}\nonumber\\
    &+\Gamma_{G}\,\delta_{1,i} \delta_{1,j}.
    \label{eq:EOM-2pt}
\end{align}
Interestingly, even though the master equation in Eq.~\eqref{eq:lindblad} involves a quartic dissipator, the equation of motion for two-point correlators closes in itself \cite{PEDSep2020}. The above equation can be re-written as
\begin{align}
    \dv{C}{t}=-{\rm i}[h_S, C]-\{C,D\}+P,
    \label{eq:EOM-2pt_matrix}
\end{align}
where $h_S$ is the single particle Hamiltonian corresponding to Eq.~\eqref{eq:system_hamiltonian_non_int} and  \begin{eqnarray}
    D_{i,j} &=& \frac{1}{2}\left(\Gamma_d \, \delta_{i,j}+ \Gamma_G \, \delta_{1,i} \delta_{1,j}\right), \nonumber \\
     P_{i,j} &=& \Gd \,C_{i,i} \, \delta_{i,j} + \GG \,\delta_{1,i} \, \delta_{1,j}.
     \label{DPcombined}
\end{eqnarray}

\subsection*{Numerical Results}
\begin{figure}[h]
    \centering
    \includegraphics[width=1.0\linewidth]{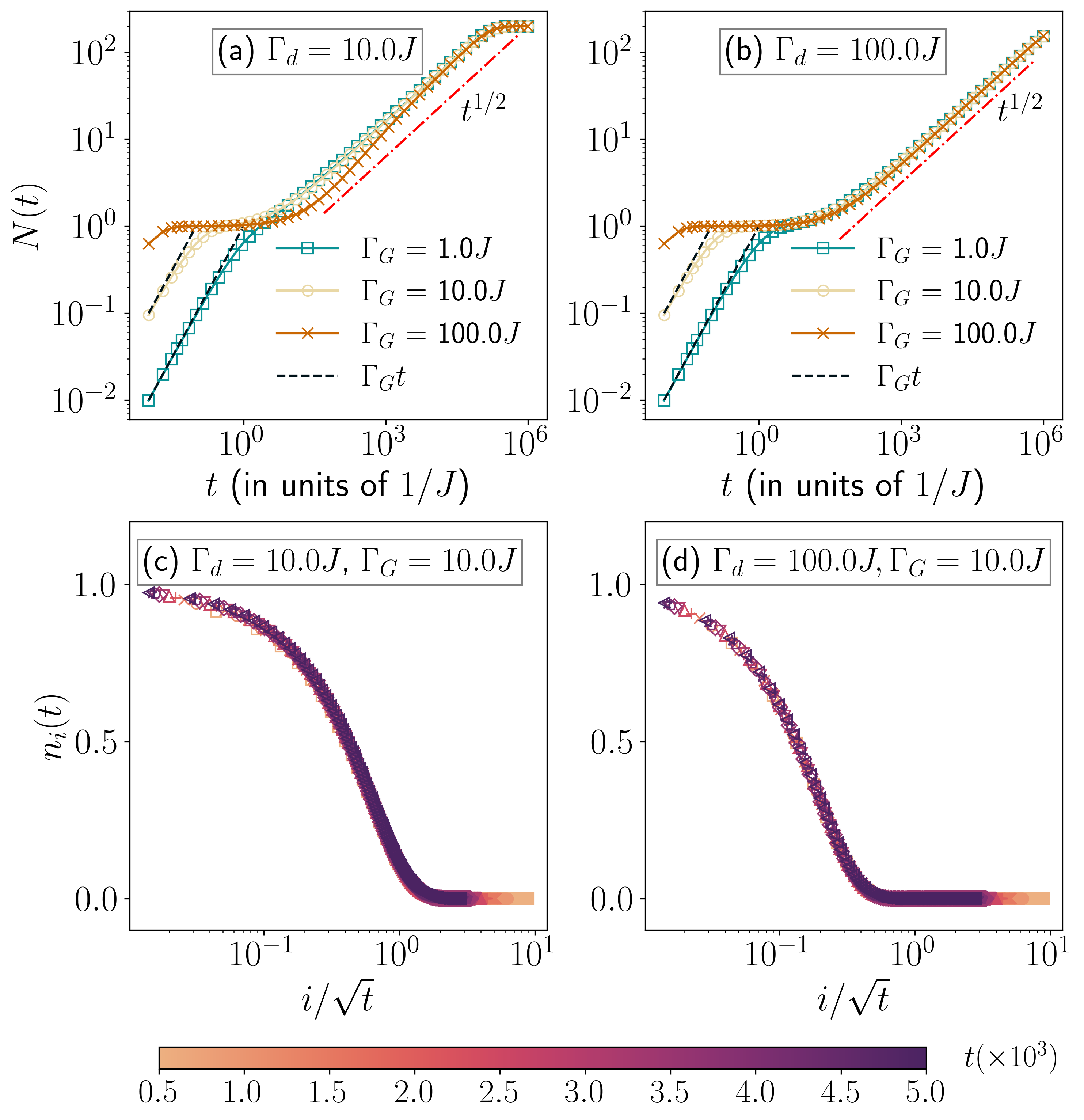}
    \caption{Quantum dynamics via exact numerics for total number of particles $N(t)$ with time $t$ (in units of $1/J$) when particles are injected from one end of the lattice and the non-interacting lattice given in Eq.~\eqref{eq:system_hamiltonian_non_int} is subjected to bulk dephasing. For (a) $\Gd = 10 J$ and for (b) $\Gd = 100 J$, with results shown for different values of $\GG$. Here, we take $L = 200$ and the hopping amplitude as $J=1$. The growth of $N(t)$ shows a crossover from linear growth ($N(t)\propto t$) to sub-diffusive growth ($N(t) \propto t^{\nu}$ with $\nu<1/2$) and finally diffusive ($N(t)\propto t^{1/2}$) before saturating to a system-size dependent value. In (a), note that when $\Gamma_G \gtrsim \Gamma_d$, the diffusive regime ($t^{1/2}$) appears at a later stage in time, as opposed to the opposite case, $\Gamma_G \lesssim \Gamma_d$.
    The local density profile $n_i(t)$ is plotted as a function of scaled lattice coordinate $i/\sqrt{t}$ at different time instances in the regime denoted by the dashed red line in (a) and (b), for (c) $\Gd = 10 J$ and (d) $\Gd = 100 J$, respectively, and $\GG = 10 J$. The plots show a perfect collapse indicating diffusive scaling in the long-time limit.}
    \label{fig:full_corr_data}
\end{figure}

In Fig.~\ref{fig:full_corr_data}, we plot the  dynamics for total number of particles 
\begin{equation}
N(t) = \sum_{i = 1}^L C_{i,i}(t)
\end{equation}
by numerically solving Eq.~\eqref{eq:EOM-2pt_matrix} along with Eq.~\eqref{DPcombined}. The initial condition is always chosen to be a vacuum state for the lattice. We show results 
for dephasing rates (a) $\Gd = 10 \, J$ and (b) $\Gd = 100 \, J$ with different injection rates $\GG =  J, 10 J,$ and $100 J$, for a fixed system size $L = 200$. We find three distinct time regimes for the growth of $N(t)$ before final equilibration takes place:  (i) First, $N(t)$ shows a linear growth in time $N(t) = \GG \, t$ up to $t \sim \mathcal{O}(1/\GG)$. (ii) This is followed by a region of sub-diffusive growth where $N(t) \sim t^\nu$ with the exponent $\nu$ continuously varying but always remaining less than $1/2$. (iii) This sub-diffusive regime is followed by a region of diffusive growth $N(t) \sim t^{1/2}$. Given the finite system size $L$, $N(t)$ eventually saturates to the asymptotic value $N_{\rm eq}=L$.

In the first regime, i.e., up to $t \sim \mathcal{O}(1/\GG)$, only the first site gets mainly populated $N(t) \sim 1$. Further flow of particles into the next lattice site is governed by the hopping strength $J$. When $\GG$, $\Gd \gg J$, the effective hopping strength, in comparison to the other time-scales in the system, decreases. This leads to the delay in the filling of the next site, which eventually leads to the regime of sub-diffusive growth. When $\GG, \Gd \sim J$, we find this sub-diffusive regime to be shorter. This is clearly seen in Fig.~\ref{fig:full_corr_data}(a) for $\GG = 1.0J$ and $\Gd = 10.0J$. 
The sub-diffusive regime lasts till $t \sim \mathcal{O}[\text{max}(\GG,\Gd)/J^2]$, after which $N(t)$ grows diffusively. This diffusive behavior is also captured in Fig.~\ref{fig:full_corr_data}(c) and (d), where we plot the local density profile $n_i(t)$ as a function of scaled lattice coordinate $i/\sqrt{t}$, at different time instances in the diffusive regime. We find a perfect collapse of the density profile.  Interestingly, we find that in the diffusive regime discussed above, the diffusion constant seems to be independent of $\GG$. This is indicated in Fig.~\ref{fig:full_corr_data} (a) and (b) by the overlapping $N(t)$ for different values of $\GG$.

For $\Gd = 10.0J$ and $\GG = 100.0J$ [Fig.~\ref{fig:full_corr_data} (a)], the dynamics takes longer to settle into the diffusive regime. This is due to the fact that $\GG>\Gd$, and the relevant timescale is now dictated by $\GG$. For $\Gd = 100.0J$, which is now the largest timescale in the system, the onset of the diffusive regime is the same for all $\Gamma_G$. Therefore, even for $\Gamma_G = 100.0J$ we do not see similar behaviour as that observed in Fig.~\ref{fig:full_corr_data} (a).

To get further insight about these different regimes, in what follows, we investigate the dynamics analytically (i) at early times and (ii) for long times by employing adiabatic approximation, valid in the strong dephasing limit.

\subsection*{Analytical results}
\noindent \textit{Early-time behaviour.--} We first discuss the prediction for $N(t)$ that follows from Eq.~\eqref{eq:EOM-2pt_matrix} at early time $t\ll 1/J$. At this timescale, it is natural to expect that only the first site which is coupled to the injection channel gets populated and coherences ($i \neq j$) among the sites are  not yet generated due to lack of particle hopping. As a result it is evident from Eq.~\eqref{eq:EOM-2pt_matrix} that the dephasing mechanism also does not get activated. Therefore, the equation of motion involving only the first site becomes,
\begin{align}
    \frac{dC_{1,1}}{dt} = \Gamma_G \, (1-C_{1,1})\quad \mathrm{for}\quad t\ll \frac{1}{J}\, .
    \label{eq:early_time_eom_C11}
\end{align}
Solving Eq.~\eqref{eq:early_time_eom_C11} with the initial condition that the lattice is initially empty, i.e., 
\begin{equation}
\label{eq:init}
C_{i,i}(t=0)=0, \quad \mathrm{for} \quad i=1,2, \cdots L, 
\end{equation}
we obtain,
\begin{align}
    C_{1,1}(t) = 1-e^{-\Gamma_G t} = \Gamma_G \, t + \mathcal{O}(t^2),
\end{align}
which matches with our numerical results, as shown in Fig.~\ref{fig:full_corr_data}(a) and (b). 

We next analytically obtain the long time behaviour for $N(t)$ which shows a diffusive exponent. For that purpose, we employ the adiabatic approximation which is accurate in the strong dephasing limit. \\
\noindent

\noindent \textit{Adiabatic approximation in the strong dephasing limit.--}
To further analyze the behavior of the system analytically for any time $t$, we consider the evolution of the particle density with the adiabatic approximation \cite{DPJul2012,DPNov2013,BSApr2015,CTMar2017,CMD2023,LTMar2025}, where we take the large dephasing limit $\Gamma_d \gg J$. Employing this limit, one can eliminate the coherence elements $C_{i,j\neq i}$ adiabatically in Eq.~\eqref{eq:EOM-2pt} and find an effective equation for the diagonal elements i.e., for local densities $n_i = C_{i,i}$. The adiabatic procedure involves the following:
We consider the equations for the coherences with the assumption that they vary slowly, i.e., $\dot{C}_{i,j}\ll \Gamma_d \, C_{i,j}$. This, from  Eq.~\eqref{eq:EOM-2pt}, leads to a relation involving $C_{i,j\neq i}$, and is given as
\begin{align}
    C_{i,j\neq i} &= -\frac{{\rm i} J}{\frac{\Gamma_G}{2}(\delta_{1,i}+\delta_{1,j})+\Gamma_d}(C_{i-1,j}\nonumber\\
    &\hspace{1.cm}+ C_{i+1,j} - C_{i,j-1} - C_{i,j+1}).
    \label{eq:c_mn_adiabatic}
\end{align}
Following Eq.~\eqref{eq:EOM-2pt}, the exact equation of motion for the local densities is given by
\begin{align}
    \dv{C_{i,i}}{t} &= -{\rm i} J (C_{i-1,i} + C_{i+1,i} - C_{i,i-1} - C_{i,i+1})\nonumber\\
    &\hspace{2cm}- \Gamma_G\,\delta_{1,i} C_{i,i} + \Gamma_G\,\delta_{1,i}.
    \label{eq:c_mm}
\end{align}
Now, using Eq.~\eqref{eq:c_mn_adiabatic} in Eq.~\eqref{eq:c_mm} and ignoring those $C_{i,j}$ terms where $|i-j|>1$ (i.e., neglecting higher order coherence), we get the equation of motion for the diagonal terms, which can be expressed as
\begin{equation}
    \dv{\Cd}{t} = \mathbb{A} \, \Cd + \mathbb{P},
    \label{eq:c_mm_eom_adiabatic}
\end{equation}
where
\begin{align}
\label{eq:Amat}
    \mathbb{A}  &= 
    \scalebox{0.95}{$
    \begin{pmatrix}
        -\alpha_2 - \GG & \alpha_2 & 0 &0 &\cdots &0&0\\
        \alpha_2 & -\alpha_1-\alpha_2 &\alpha_1& 0&\cdots&0&0\\
        0 & \alpha_1 & -2\alpha_1 & \alpha_1 & \cdots &0&0\\
        \vdots &&\ddots&\ddots&\ddots&\vdots&\vdots\\
        0&&\cdots&&\alpha_1&-2\alpha_1&\alpha_1\\
        0&&\cdots&&&\alpha_1&-\alpha_1
    \end{pmatrix}
    $}
\end{align}
and 
\begin{equation}
\mathbb{P} = [\Gamma_G, 0 ,\cdots,0].
\label{eq:P}
\end{equation}
Here $\Cd(t)$ is a column vector consisting of diagonal elements (local density) of the matrix $C$. In Eq.~\eqref{eq:Amat}, we define 
\begin{eqnarray}
\alpha_1 = \frac{2J^2}{\Gd},\quad\quad
\alpha_2 = \frac{2J^2}{\big(\frac{\GG}{2}+\Gd\big)}. 
\label{eq:alpha12}
\end{eqnarray}
The solution for Eq.~\eqref{eq:c_mm_eom_adiabatic}, along with the empty initial condition [Eq:~\eqref{eq:init}] is given by
\begin{equation}
    \Cd(t) = \left(e^{\mathbb{A}t}-\mathbb{I}\right)\mathbb{A}^{-1}\,\mathbb{P}\,,\label{eq:c_diag_adiabatic_solution}
\end{equation}
where $\mathbb{I}$ is $L\times L$ identity matrix. Note that Eq.~\eqref{eq:c_diag_adiabatic_solution} albeit exact, is difficult to compute analytically. To get analytical solutions, the matrix in Eq.~\eqref{eq:Amat} needs to be simplified further. Interestingly, it turns out based on the numerical observation that the diffusive regime always remains robust, independent of the value of $\GG$, for large system sizes. Therefore, to analytically evaluate $N(t)$ and the diffusion constant,  we further simplify our calculations by taking $\GG = \alpha_1$, and since in the adiabatic approximation, $\Gd \gg J$, we have $\alpha_2 \approx \alpha_1$ from Eq.~\eqref{eq:alpha12}.
\begin{figure*}
    \centering
    \includegraphics[width=1\linewidth]{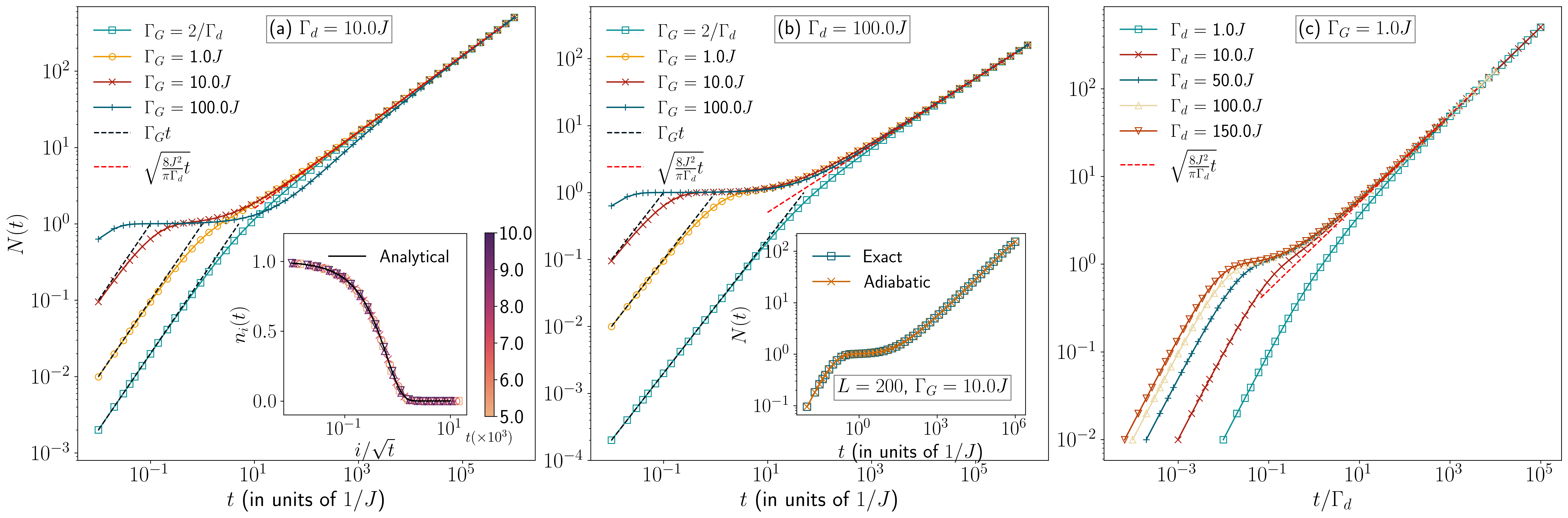}
    \caption{The dynamics of total number of particles $N(t)$ in the adiabatic limit with time $t$ (in units of $1/J$) when particles are injected. For (a) $\Gd = 10.0J$ and (b) $\Gd = 100.0J$, for different values of $\GG$ and $J = 1$.
    $N(t)$ shows a crossover from ballistic ($N(t)\propto t$), to sub-diffusive ($N(t) \propto t^{\nu}$ with $\nu<1/2$) and finally to diffusive ($N(t)\propto t^{1/2}$). The density profile $n_i(t)$ is plotted as a function of $i/\sqrt{t}$ at different time instances in the inset of (a) for $\Gd = 10.0J$ and $\GG = 2/\Gamma_d$, showing agreement with the analytical result obtained in Eq.~\eqref{eq:n_xt_adiabatic_special_case}. In the inset of (b) $N(t)$ is plotted using exact numerics [Eq.~\eqref{eq:EOM-2pt_matrix}] and adiabatic approximation [Eq.~\eqref{eq:c_diag_adiabatic_solution}] for $L = 200, \Gd = 100.0J$, and $\GG = 10.0J$, showing good agreement between the two. (c) $N(t)$ is plotted as a function of scaled time axis $t/\Gamma_d$ for $\GG = 1.0J$, and different values of $\Gd$. The system size is fixed to $L = 1000$ for all cases, unless otherwise specified.}
    \label{fig:adiabatic}
\end{figure*}
As a result, the propagator $\mathbb{A}$ simplifies to
\begin{align}
    \mathbb{A} &= \alpha_1\begin{pmatrix}
        -2  & 1 & 0 &0 &\cdots &0&0\\
        1 & -2 &1& 0&\cdots&0&0\\
        0 & 1 & -2 & 1 & \cdots &0&0\\
        \vdots &&\ddots&\ddots&\ddots&\vdots&\vdots\\
        0&&\cdots&&1&-2&1\\
        0&&\cdots&&&1&-1
    \end{pmatrix}.\label{eq:weak_gain_A}
\end{align}
The eigenvalues and eigenvectors of the above matrix in Eq.~\eqref{eq:weak_gain_A} can be easily obtained and are given by \cite{yueh2005eigenvalues},
\begin{align}
    &\lambda_k = -2\alpha_1\Bigg\{1-\cos{\Bigg[\frac{(2k-1)\pi}{2L+1}\Bigg]}\Bigg\}, \quad k\in[1,L]\\
    & u_j^k = \frac{2}{\sqrt{2L+1}}\sin{\Bigg[\frac{(2k-1)j\pi}{2L+1}\Bigg]}. \quad j\in [1,L]
\end{align}
The matrix $\mathbb{A}$ can be diagonalized as $\mathbb{A}_d=U^T\mathbb{A}U$, where $U = [\ket{u^1}, \ket{u^2}, \cdots, \ket{u^L}]$. Using Eq.~\eqref{eq:c_diag_adiabatic_solution}, the expression of local density $n_i(t)$ can then be obtained as,
\begin{align}
    n_i(t) &= \frac{4\GG}{2L+1}\sum_{k=1}^L\frac{e^{\lambda_k t}-1}{\lambda_k}\nonumber\\
    &\hspace{1.5cm}\times\sin{\left[\frac{(2k-1)\pi}{2L+1}\right]}\sin\left[\frac{(2k-1)i\pi}{2L+1}\right] .
    \label{eq:n_i_exact_adiabatic_special_case}
\end{align}
Next, we consider the large system size limit, i.e., $L \rightarrow\infty$. We set $k\pi/(2L+1) \coloneqq \Tilde{k}$. In the limit $L \to \infty$, we can shift from a discrete sum to a continuum integral, $1/(2L+1)\sum_{k = 1}^L \to (1/\pi)\int_0^{\pi/2}\dd \Tilde{k}$. The expression for local densities in Eq.~\eqref{eq:n_i_exact_adiabatic_special_case} becomes
\begin{align}
    n(x,t) &=\frac{4\GG}{\pi}\int_0^{\pi/2}\dd\Tilde{k}\,\frac{1-e^{-4\alpha_1 t \sin^2\Tilde{k}}}{4\alpha_1 \sin^2\Tilde{k}}\nonumber\\
    &\times\sin(2\Tilde{k})\sin(2\Tilde{k}x),
    \label{nx_int}
\end{align}
where we have set the lattice spacing to $1$. Now, considering the large time limit, i.e., $t\gg 1/J$, we obtain a compact expression for the local density profile as
\begin{align}
    n(x,t) &= 1-{\rm Erf}\left(\frac{x}{\sqrt{4\alpha_1 t}}\right),
\label{eq:n_xt_adiabatic_special_case}
\end{align}
where ${\rm Erf}(z)$ is the error function. To obtain Eq.~\eqref{eq:n_xt_adiabatic_special_case}, we have used the fact that at large times, only small $\Tilde{k}$ contributes to the integral in Eq.~\eqref{nx_int} as the higher modes are exponentially suppressed and $\lim_{\Tilde{k}\to 0}\sin \Tilde{k} \approx \Tilde{k}$. Hence, in this regime, we can also change the upper limit of the integral in Eq.~\eqref{nx_int} from $\pi/2$ to $\infty$, which subsequently yields Eq.~\eqref{eq:n_xt_adiabatic_special_case}. The total number of particles at long times is given by
\begin{align}
    N(t) &= \int_0^{\infty}\dd x \, n(x,t) = \sqrt\frac{8J^2 t}{\pi\Gd},    \label{eq:Nt_adiabatic_special_case_long_time}
\end{align}
which gives the diffusive scaling exponent $(t^{1/2})$ with the diffusion constant, 
\begin{align}
    D = \sqrt{\frac{8J^2}{\pi \Gd}}.
    \label{eq:diffusion_constant_non_int}
\end{align}
In Fig.~\ref{fig:adiabatic} (a) and (b), we plot $N(t)$ as a function of time (in units of $1/J$) for $\Gd = 10.0J$ and $\Gd = 100.0J$, respectively, for different values of $\GG$ using the adiabatic approximation i.e., by solving Eq.~\eqref{eq:c_diag_adiabatic_solution}. We consider the system size $L  =1000$. We find that the emergence of the diffusive regime in the long-time limit and the associated diffusion coefficient always remains the same, independent of the value of $\GG$. The value exactly matches the analytical prediction, as given in Eq.~\eqref{eq:diffusion_constant_non_int}. In the inset of Fig.~\ref{fig:adiabatic} (a), we plot the density profiles at different time instances as a function of $i/\sqrt{t}$. The collapsed density profile exactly overlaps with the density profile given by Eq.~\eqref{eq:n_xt_adiabatic_special_case}, which further confirms the diffusive behavior. In the inset of Fig.~\ref{fig:adiabatic} (b), $N(t)$ is plotted using exact numerics [Eq.~\eqref{eq:EOM-2pt_matrix}] and adiabatic approximation [Eq.~\eqref{eq:c_diag_adiabatic_solution}] for $L = 200, \Gd = 100.0J$, and $\GG = 10.0J$, showing good agreement between the two. In Fig.~\ref{fig:adiabatic} (c) we plot $N(t)$ as a function of scaled time $t/\Gd$, for $\GG = 1.0J$, and for different values of $\Gd \geq \GG$. We find that the time scale for the onset of diffusive behavior collapses for all values of $\Gd$, except for $\Gd = 1.0J$. This indicates that the sub-diffusive regime lasts from $t \sim O(1/\GG)$ to $t\sim O(\Gd/J^2)$, when $\Gd > \GG$.

\begin{figure}
    \centering
    \includegraphics[width=0.8\linewidth]{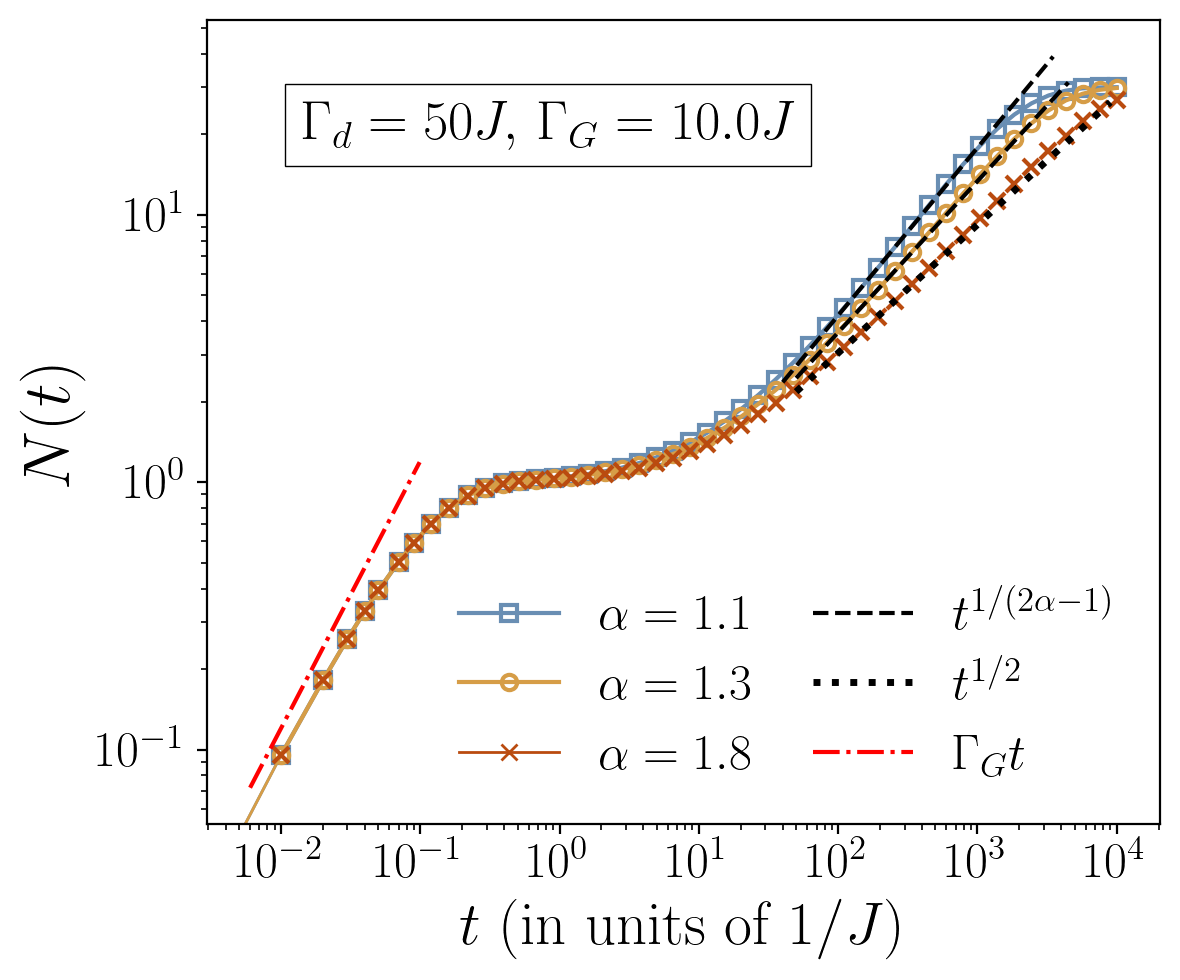}
    \caption{The dynamics of total number of particles $N(t)$ with time $t$ for the long range lattice system, defined in Eq.~\eqref{eq:HLR}, when the particles are injected from the left. For different long-range hopping exponent $\alpha$, the late time dynamics is different. For $1<\alpha<1.5$, the dynamics is superdiffusive with time dependence $t^{1/2\alpha-1}$ (black dashed line) and for $\alpha>1.5$ (black dotted line), the dynamics is diffusive with $t^{1/2}$ scaling.}
    \label{fig:LR}
\end{figure}

\section{Generalization to long-ranged hopping models}
\label{sec:longr}

In the previous section [Sec.~\ref{sec:setupI}] we discussed the case when the Hamiltonian is a short-ranged tight binding model [Eq.~\eqref{eq:system_hamiltonian_non_int}]. We now briefly discuss the case where the underlying model is long-ranged with hopping between sites of  power-law form. More precisely, the analog of Eq.~\eqref{eq:system_hamiltonian_non_int} is taken to be
\begin{equation}
H_S=-J \sum_{m=1}^{L} \sum_{i=1}^{L-m}\frac{c^{\dagger}_{i}c_{i+m}}{m^\alpha}+{\rm h.c.},
\label{eq:HLR}
\end{equation} 
where $\alpha$ characterizes the long-range hopping exponent. The rest of the setup, i.e., bulk dephasing and a localized source remains the same [Eq.~\eqref{eq:lindblad}], as in Sec.~\ref{sec:setupI}. Long-ranged systems [Eq.~\eqref{eq:HLR}] are known to exhibit interesting anomalous behavior as a function of the exponent $\alpha$~\cite{ASJan2020,ADAug2024}. For example, in Ref.~\onlinecite{ADAug2024}, the authors examine a 1D fermionic lattice with long-range hopping and dephasing noise. They find that for $1<\alpha<1.5$, transport is superdiffusive, while for $\alpha>1.5$ it becomes diffusive.  The work highlights how long-range interactions and dephasing lead to unconventional transport in open quantum systems. However, when such long-ranged systems with bulk dephasing are subjected to a localized source, the quantum dynamics of total number of particles $N(t)$ remains far from obvious. 

In Fig.~\ref{fig:LR}, we plot the time dynamics of $N(t)$ for different values of $\alpha$. The growth of $N(t)$ at short-time is once again linear in $t$ with the same growth rate $2 \Gamma_G$. A sub-diffusive plateau once again is observed, after which a clear $\alpha$ dependent power-law growth is obtained. More precisely, we find 
\begin{equation}
N(t) \sim 
\begin{cases}
t^{\frac{1}{2\alpha-1}} \quad \mathrm{for} \,\,\,\, 1 < \alpha <3/2, \\
t^{\frac{1}{2}} \quad \mathrm{for} \quad  \alpha \geq 3/2,
\end{cases}
\end{equation}
which confirms anomalous superdiffusive growth in the effectively long-range regime i.e., for $1<\alpha<3/2$. For $\alpha>3/2$, the system effectively falls into the universality class of short-range models, exhibiting diffusive growth.

Having discussed the problem of injection in a dephased non-interacting lattice, both short [Sec.~\ref{sec:setupI}] and long-ranged [Sec.~\ref{sec:longr}], a natural question is what happens when injection is performed in an inherently many-body interacting setup. This is what is discussed next. 

\section{Setup and results for the interacting lattice with dephasing}
\label{sec:inherently_interacting}

In this section, we discuss the case of an interacting fermionic lattice systems, with particles being injected from one end of the lattice, and the dephasing mechanism is active at all sites of the lattice. Furthermore, if we add on-site potentials, we can map the setup to a closed $XXZ$ spin$-1/2$ chain. The Hamiltonian of the closed setup in the fermionic language is
\begin{align}
    H =& \frac{J}{2}\sum_{i = 1}^{L-1}(c_i^\dagger c_{i+1} + c_{i+1}^\dagger c_i + \Delta n_i n_{i+1}) \nonumber \\ &- J\Delta\sum_{i = 1}^L n_i + \frac{J\Delta}{2}(n_1 + n_L) + \frac{J\Delta}{2}(L-1).
    \label{eq:int_fermions_hamiltonian}
\end{align}
Using the Jordan-Wigner transformation, such a setup maps to the isolated XXZ spin$-1/2$ chain, whose Hamiltonian is given by,
\begin{align}
    H_{\rm XXZ} &= J\sum_{i=1}^{L-1}\left(S^x_{i}S^x_{i+1} + S^y_{i}S^y_{i+1} + \Delta S^z_{i}S^z_{i+1}\right),
    \label{eq:XXZ_hamil}
\end{align}
where $S_{i}^{x,y,z} = \sigma_{i}^{x,y,z}/2$ with $\sigma_{i}^{x,y,z}$ being the Pauli matrices for site $i$.  $\Delta$ is the $z$ anisotropy. In the particle picture, $\Delta$ is the many-body interacting strength, as can be seen from Eq.~\eqref{eq:int_fermions_hamiltonian}. The above setup is subjected to (i) the injection of magnetization (analogous to particles in fermionic language) from one end and  (ii) dephasing at every site. These processes are achieved by the following Lindblad jump operators,
\begin{align}
    &L_1 = \sqrt{\Gamma_G}\, S^+_1, \quad \mathrm{and}
    &L_{i}=\sqrt{\Gamma_d}\,S^{z}_i, \quad \mathrm{for}\,\, i=1,2, \cdots L
\end{align}
where $\Gamma_G$ and $\Gd$ are the injection and dephasing rate respectively. Here $S^{\pm}_i = S^x_i + i \, S^y_i $. The equation of motion for the density matrix $\rho$ of is modeled by the GKSL quantum master equation
\begin{align}
    \dot{\rho} = - {\rm i}[H_{\rm XXZ},\rho]&+\Gamma_G\left[S^+_1\rho S^-_1 - \frac{1}{2}\left\{S^-_1 S^+_1,\rho\right\}\right]\nonumber\\&+\Gamma_d\sum_{i=1}^L\left[S^z_i\rho S^z_i - \rho\right].
    \label{eq:QME}
\end{align}
We solve the dynamics of this setup using the TEBD algorithm starting with an initial state where all the sites are in the down-polarised state, i.e., 
\begin{equation}
  \rho(0)= \ket{\psi(0)} \bra{\psi(0)} = \ket{\dn \dots \dn} \bra{\dn \dots \dn}.
\end{equation}
In the particle picture, this corresponds to a state where the lattice is empty. Eq.~\eqref{eq:QME}, can be written in a vectorized form as,
\begin{equation}
    \ket{\dot{\rho}} = \mathcal{L}\ket{\rho},
\end{equation}
where we vectorize the density matrix by column stacking, and the Liouvillian superoperator is given by
\begin{align}
    \mathcal{L} &= -{\rm  i}\left(\mathbb{I}\otimes H_{\rm XXZ} - H_{\rm XXZ}^T\otimes \mathbb{I}\right)\nonumber\\
    &+ \! \Gamma_G \left[(S^-_1)^T\otimes S^+_1 - \frac{1}{2}\left[ \mathbb{I}\otimes (S^-_1S^+_1) + (S^-_1S^+_1)^T\otimes \mathbb{I} \right]\right], \nonumber\\
    &+ \!\Gd\sum_{i=1}^L\left[S^z_i\otimes S^z_i - \mathbb{I}\right].
    \label{eq:vec_liov}
\end{align}
The state at any time $t$ can be then be obtained as 
\begin{equation}
    \ket{\rho(t)} = e^{\mathcal{L}t}\ket{\rho(0)}.
    \label{eq:rho_t}
\end{equation}

\begin{figure*}
    \centering
    \includegraphics[width=1.0\linewidth]{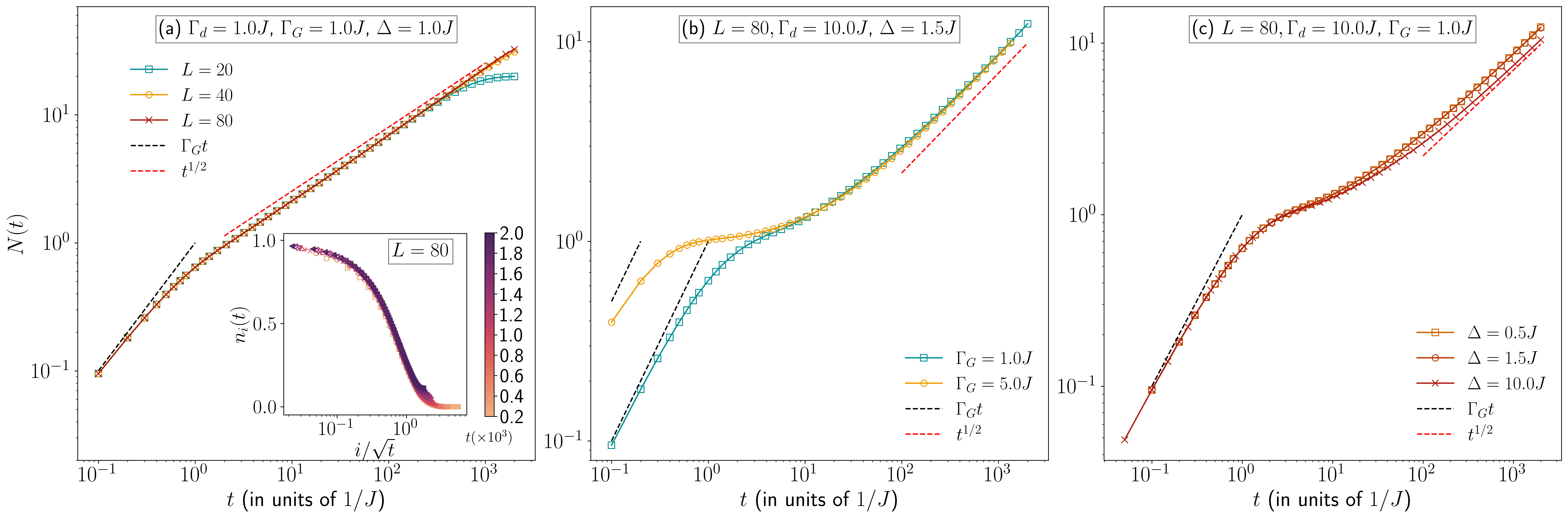}
    \caption{Plot for quantum dynamics of average total particle number $N(t)$ with time $t$ (in units of $1/J$) for the XXZ lattice given in Eq.~\eqref{eq:XXZ_hamil}. (a) Growth of $N(t)$ is plotted for different system sizes $L=20, 40$, and $80$. The other parameters are given as $\Gamma_d=1.0J$, $\Gamma_G=1.0J$, and $\Delta=1.0J$ (isotropic case). Initially, $N(t)$ grows linearly as $\Gamma_G t$ until the dephasing effect emerges. At late times, $N(t)$ grows diffusively. In the inset, the density profile $n_i$ is plotted with respect to the scaled lattice site $i/\sqrt{t}$ for different time instants and for system size $L=80$. A perfect scaling collapse of $n_i$ data is observed for different time instants, confirming the late-time diffusive dynamics. (b) Plot for $N(t)$ with time $t$ for different $\Gamma_G$ values with $\Gamma_d=10.0J$, $\Delta=1.5J$. We observe that, though the initial dynamics are sensitive to the value of $\Gamma_G$ because initially $N(t)\sim \Gamma_G t$ (dashed black line), the late-time dynamics are insensitive to the value of $\Gamma_G$. In (c), we have shown that the diffusion constant depends on $\Delta$. We fix the parameters to be $L = 80, \Gd = 10.0J$, and $\GG = 1.0J$. We have fixed the bond dimension at $\chi = 200$ in all the figures.}
    \label{fig:xxz_plot}
\end{figure*}

\subsection{Numerical results}
\label{sec:interacting_numerics}
In this subsection, we present numerical results for the quantum dynamics of lattice filling in the interacting model described in Eq.~\eqref{eq:XXZ_hamil}.
To simulate the dynamics governed by Eq.~\eqref{eq:vec_liov} and Eq.~\eqref{eq:rho_t}, we employ the TEBD algorithm adapted for open systems~\cite{MZNov2004}. The implementation details are provided in Appendix~\ref{app:TEBD}. For the simulations, we use the maximum bond dimension $\chi=200$ \footnote{We have seen that this is enough to obtain convergence for all cases.}, and we verify the convergence of our results against larger bond dimensions. The time evolution is performed using a fourth-order Trotter decomposition, which introduces an error $\mathcal{O}(\delta t^{5})$ at each time step (see Appendix~\ref{app:TEBD} for details). We fix the time step to $\delta t=0.05$ (in units of $1/J$). We compute total particle number $N(t)=\sum_{i=1}^{L} \langle S^z_i(t)\rangle+\frac{L}{2}$ for different system sizes $L$,  anisotropy parameter values $\Delta$, and injection rate $\Gamma_G$. The hopping strength is fixed to unity ($J=1$) in all cases.

In Fig.~\ref{fig:xxz_plot}, we have presented our results.  In Fig.~\ref{fig:xxz_plot} (a), the particle number $N(t)$ is plotted as a function of time $t$ (in units of $1/J$) for different system sizes $L=20, 40, 80$. One can observe that, similar to the non-interacting lattice setup, even in the interacting case, the initial growth of $N(t)$ is linear with a rate $\Gamma_G$. At this timescale, only the first site gets populated by the local injection, and the system remains in a single particle sector. Therefore, the effect of many-body interaction does not emerge, and thus growth remains linear in time $t$, as predicted for the non-interacting setup. In fact,  the dynamics is governed by the equation of motion of the first site as given in Eq.~\eqref{eq:early_time_eom_C11}. Such linear growth continues till $t\sim O(1/\Gamma_G)$ when the first site acquires sizable filling.

On the other hand, the long-time growth of $N(t)$, in all cases, shows a diffusive $\sqrt{t}$ scaling before the saturation effect due to finite size kicks in. The emergence of diffusive scaling is also confirmed by the local density profile plot in the inset of Fig.~\ref{fig:xxz_plot} (a), where for different time instants, the local density profile $n_i(t)$ is plotted against rescaled coordinate $i/\sqrt{t}$. A perfect collapse of $n_i(t)$ for different time snapshots is observed. Therefore, one can conclude that, in the presence of a bulk dephasing mechanism, the late-time filling dynamics are universal and the growth is diffusive as it emerges both in non-interacting and interacting lattice setups.

Similar to Secs.~\ref{sec:setupI} and \ref{sec:longr}, there is an intermediate regime, characterized by continuously varying time exponents. The window of such a regime crucially depends on the choice of parameters. The onset of the diffusive regime is governed by $\Gd, \GG,$ and $\Delta$. However, when all these parameters of $O(J)$, we do not see any sub-diffusive plateau, as is evident in Fig.~\ref{fig:xxz_plot} (a) where $\Gamma_G=\Gamma_d=\Delta=1.0J$. Interestingly, this is not the case when $\GG$ and $\Gd$ are different, for example, in Fig.~\ref{fig:xxz_plot}(b), similar to the non-interacting setup, the ballistic to diffusive crossover happens through an intermediate sub-diffusive plateau. For the chosen parameters in Fig.~\ref{fig:xxz_plot}(b), the time scale of the sub-diffusive to diffusive crossover is of $\mathcal{O}(\Gamma_d)$. Such a sudden slowing down of the lattice filling dynamics is rooted in the fact that the interplay of injection and dephasing effects leads to a reduced effective hopping strength between lattice sites by enabling various incoherent scattering mechanisms.

Next, for the diffusive regime, we comment on the diffusion constant following Fig.~\ref{fig:xxz_plot}(b). We observe that when $\Gamma_d$ is large compared to all the other parameters of the setup, 
interestingly, the diffusion constant appears to be insensitive and is solely determined by the dephasing strength. This is analogous to the non-interacting scenario. 
In Fig.~\ref{fig:xxz_plot}(b), we plot $N(t)$ for different values of injection strength $\Gamma_G$. Though the initial linear growth is sensitive to the $\Gamma_G$ value, the late-time diffusive dynamics are completely insensitive to it. Not only that, the crossover to diffusive dynamics from the intermediate sub-diffusive regime is also independent of the value of $\Gamma_G$, when $\Gd > \GG$. However, similar to the non-interacting cases, this crossover time is governed by the most dominant parameter in the system, as it will be made clear in Fig.~\ref{fig:xxz_plot} (c).

In Fig.~\ref{fig:xxz_plot} (c) we plot $N(t)$ for different values of $\Delta$, keeping $\Gd = 10.0J$ and $\GG = 1.0J$. In this case, we can clearly observe that the diffusion constant depends on $\Delta$, which is manifest in the fact that the curves do not overlap in the diffusive regime. Although this dependence is not noticeable for $\Delta = 0.5J$ and $\Delta = 1.5J$, for large enough $\Delta$, it becomes clear.

\section{Summary}
\label{sec:summary}
In summary, we have investigated the quantum dynamics of lattice filling for two classes of systems, i) non-interacting (nearest-neighbor and long-ranged), and ii) interacting (nearest-neighbor), with both being subjected to local dephasing probes at each site. We show that in the initial short-time dynamics, the particle (net magnetization) growth is linear in time with a rate proportional to the injection strength. At late times, the growth for both cases becomes diffusive. The corresponding local density profile also shows a diffusive scaling collapse. Interestingly, in both cases, the diffusion coefficient is independent of the injection rate. For the non-interacting nearest-neighbor case, we employ the adiabatic approximation in the large dephasing limit, to analytically compute the evolution of the density profile and the total number of particles at late times. We explicitly show that the diffusion constant is indeed only a function of $\Gd$. For the interacting system, we show that the diffusion constant is again independent of $\GG$, and depends on $\Gd$ and the interaction strength $\Delta$. At intermediate time, the growth shows a plateau-like dynamical regime with continuously changing sub-diffusive exponent. Such slow sub-diffusive dynamics is a combined effect of local injection and dephasing, which reduces the effective hopping amplitude to the neighbouring site. We show that the onset of the diffusive regime is governed by the most dominant parameter in the system ($\Gd, \GG$, and $\Delta$). For cases when the parameters are $\sim O(J)$, this sub-diffusive plateau does not exist. Furthermore, for the long-ranged non-interacting model, we find that after the initial linear growth, followed by a sub-diffusive plateau, the total number of particles grows super-diffusively for $1<\alpha<3/2$ and diffusively for $\alpha\geq 3/2$.

We believe that our study is relevant in the context of filling major gaps in the literature on particle injection in lattices with environmental effects, which induce inelastic scattering and phase randomization processes, and interactions. The problem of particle injection has been studied in several setups. Similarly, the role of dephasing in non-interacting and interacting models has been investigated. However, a natural question to ask in this context is how the interplay of these different mechanisms influences the quantum dynamics. Our work provides significant insights, both numerically and analytically, in this context. It reveals rich dynamical behavior, encompassing universal dynamical scaling and anomalous features across multiple time scales, with broad relevance to diverse quantum simulation platforms.

In the future, it will be interesting to understand the behaviour of higher-order fluctuations for particle number growth in such setups by obtaining full statistics of total particle number. The effect of correlated dephasing (by suitable generalization of the jump operators in the GKSL dynamics) in the lattice filling dynamics can also be worth exploring, as few recent studies show evidence of faster than diffusive dynamics in such cases~\cite{renji2023,schiro2025}. Therefore, it will be interesting to investigate the interplay of correlated dephasing, local injection, and many-body interaction in such setups. Another interesting avenue to explore is the bosonic counterpart of the setup that we explored in this work.

\section*{Acknowledgments}
BKA acknowledges CRG Grant No. CRG/2023/003377 from Science and Engineering Research Board (SERB), Government of India. 
KG and BKA acknowledge the National Supercomputing Mission (NSM) for providing computing resources of ‘PARAM Brahma’ at IISER Pune, which is implemented by C-DAC and supported by the Ministry of Electronics and Information Technology (MeitY) and DST, Government of India. K.G. would like to acknowledge the Prime Minister’s Research
Fellowship (ID- 0703043), Government of India for funding. BKA acknowledges the hospitality of the International Centre of Theoretical Sciences (ICTS), Bangalore, India, under the associateship program. DP acknowledges
support from the Ministry of Education Singapore under the grant MOE-T2EP50123-0017. TR and MK acknowledge support from the Department of Atomic Energy, Government of India, under project No.~RTI4001.

\appendix

\setcounter{figure}{0}
\renewcommand{\thefigure}{A\arabic{figure}}

\section{Numerical details of TEBD for open systems}
\label{app:TEBD}

In this appendix, we discuss the time-evolving block decimation (TEBD) algorithm~\cite{USJan2011} which is used in this work to numerically evolve the density matrix for the open system described by Eq.~\eqref{eq:QME} of the main text. First, we map the density operator $\rho$ to a state $\ket{\rho}$ using the Choi-Jamiolkowski isomorphism. This is done by stacking the columns of the density matrix. Using this formalism, any action of operators $\hat{A}$ and $\hat{B}$ on $\rho$ maps to the superoperator
\begin{equation}
    \hat{A}\rho\hat{B} = \hat{\mathcal{O}}\ket{\rho} = (\hat{B}^T\otimes\hat{A})\ket{\rho}.
\end{equation}
The dimension of the space in which these vectorised states belong is the square of the dimension of the Hilbert space, which in this case is $d = 4^L$. The dimension of each site becomes $4$. We represent the state of such a system as a matrix product state (MPS)~\cite{MZNov2004}
\begin{equation}
    \ket{\rho} = \sum_{j_1,j_2,\dots,j_L=1}^4B^{(1)j_1}B^{(2)j_2}\dots B^{(L)j_L}\ket{j_1,j_2,\dots,j_L},
    \label{eq:MPS_rho}
\end{equation}
where $B^{(n)j_n}$ are matrices at site $n$ and $j_n's$ are the vectorised physical indices with dimension 4. The dimension of these matrices, known as the bond dimension, grows exponentially with the system size. To circumvent this issue, we set a cut-off in the bond dimension $\chi$.

The Liouvillian superoperator $\mathcal{L}$, given by Eq.~\eqref{eq:vec_liov} of the main text, is of the form,
\begin{align}
    \mathcal{L} = \sum_{n=1}^{L-1}\mathcal{L}_{n,n+1},
\end{align}
where $\mathcal{L}_{n,n+1}$ is a two-site superoperator acting on sites $n$ and $n+1$. From Eq.~\eqref{eq:vec_liov}, the two-site superoperators are given by
\begin{widetext}
\begin{align}
    \mathcal{L}_{n,n+1} = 
        &-{\rm i} J\Big[[\In{n}\otimes S^x_n][\In{n+1}\otimes S^x_{n+1}] + [\In{n}\otimes S^y_n][\In{n+1}\otimes S^y_{n+1}] + \Delta[\In{n}\otimes S^z_n][\In{n+1}\otimes S^z_{n+1}]\nonumber\\
        & -[(S^x_n)^T\otimes\In{n}][(S^x_{n+1})^T\otimes\In{n+1}] - [(S^y_n)^T\otimes\In{n}][(S^y_{n+1})^T\otimes\In{n+1}] - \Delta[(S^z_n)^T\otimes\In{n}][(S^z_{n+1})^T\otimes\In{n+1}]\Big]\nonumber\\
        & +  \Gd [S^z_n\otimes S^z_n - \In{n}\otimes\In{n}] + \delta_{n,1}\GG\left[(S_1^+)^T\otimes S_1^+ - \frac{1}{2}[\In{1}\otimes(S^-_1S^+_1) + (S^-_1S^+_1)^T\otimes \In{1}]\right].
    \label{eq:two_site_liouv}
\end{align}
\end{widetext}
The fourth order trotterization of the propagator $\exp(\delta t\mathcal{L})$ is given as,
\begin{equation}
    e^{\delta t\mathcal{L}}=\mathcal{U}(\delta t_1)\mathcal{U}(\delta t_1)\mathcal{U}(\delta t_2)\mathcal{U}(\delta t_1)\mathcal{U}(\delta t_1)+O(\delta t^5).
\end{equation}
Here $\mathcal{U}(\delta t_i)$ is given by,
\begin{equation}
    \mathcal{U}(\delta t_i)=e^{\mathcal{L}_{\rm odd}\delta t_i /2}e^{\mathcal{L}_{\rm even}\delta t_i}e^{\mathcal{L}_{\rm odd}\delta t_i/2},
\end{equation}
where 
\begin{eqnarray}
\mathcal{L}_{\rm odd} = \sum_{\rm n \in odd} \mathcal{L}_{n,n+1}, \quad 
\mathcal{L}_{\rm even} = \sum_{\rm n \in even} \mathcal{L}_{n,n+1}, 
\end{eqnarray}
and
\begin{align}
    \delta t_1 = \frac{\delta t}{4 - 4^{1/3}},\quad
    \delta t_2 = \delta t - 4\delta t_1.
\end{align}
The vectorised density matrix at time $t = m \delta t$ is given by
\begin{align}
    \ket{\rho(t)} = [\mathcal{U}(\delta t_1)\mathcal{U}(\delta t_1)\mathcal{U}(\delta t_2)\mathcal{U}(\delta t_1)\mathcal{U}(\delta t_1)]^m\ket{\rho(0)}.
\end{align}
The expectation value of any observable $\mathcal{O}$ in the state $\rho$ can be computed using
\begin{align}
    \expval{\mathcal{O}} = \frac{\Tr[\rho \mathcal{O}]}{\Tr \rho} = \frac{\bra{\mathbb{I}} \mathbb{I}\otimes \mathcal{O}\ket{\rho}}{\bra{\mathbb{I}}\rho\rangle},
\end{align}
where $\ket{\mathbb{I}}$ is the vectorised identity matrix.

\bibliography{references}

\end{document}